\documentclass{cls}

\usepackage{amsmath}
\usepackage{subcaption}
\usepackage{svg}

\title{Turbulent Boundary Layer Height Scales in Hurricanes}

\authors{Kishore R. Sathia, \aff{a} and
Marco G. Giometto,\aff{a}\correspondingauthor{M.G. Giometto, mg3929@columbia.edu}
}
\affiliation{\aff{a} Department of Civil Engineering and Engineering Mechanics, Columbia University, New York, NY 10027, USA}
 
\abstract{
Boundary layer processes drive the air–sea exchange of momentum, heat, and moisture that powers and shapes hurricanes.
The height of the boundary layer is a critical parameter in engineering and meteorological models of hurricane wind speed, turbulence intensity, and storm strength. 
Existing models rely on a height scale derived with the assumption of a constant eddy viscosity, a strong simplification that limits physical accuracy.
This work proposes formulae for the turbulent boundary layer height in hurricanes outside the eyewall.
The proposed scalings are $u_\star/\beta$ for neutral stratification, and $u_\star/\sqrt{\beta N}$ for stable stratification, where $u_\star$ is the friction velocity, $\beta$ is the absolute fluid vorticity and $N$ is the Brunt-V\"ais\"al\"a frequency of the background stratification.
These scalings are analogous to those used in the literature for neutrally and stably stratified turbulent atmospheric boundary layers.
The formulae are backed by analytical derivation and validated against velocity profiles from large-eddy simulations and field observations. 
They are predictive to within $2.5\%$ relative error on average and yield a good collapse of the simulated and observational velocity profiles away from the surface.
The results further enable quantitative relationships between boundary layer height and other characteristic scales, including the height of maximum wind speed and the depth of the inflow layer.
The proposed expressions offer a practical basis for interpreting observational data, informing mesoscale simulations, and specifying turbulent flow statistics in wind engineering and coastal resilience.
}

\begin{document}

\maketitle

\section{Introduction}
\label{sec:Intro}
Boundary layer processes in hurricanes are of interest for applications in wind engineering, wind energy, coastal resilience and meteorology.
A key parameter in the functional form of wind profiles and turbulent statistics is the boundary layer (BL) height. 
Analytically, it is well established for the hurricane boundary layer (HBL) idealized with a constant eddy viscosity $K$ that the HBL height scales as $\sqrt{2K/I}$ where $I$ is the inertial stability parameter \citep{meng1995analytical, kepert2001dynamics, sathia2025analytical}. 

However, there remains a significant gap in the literature when attempting to connect this scaling with models that are practically applicable.
In the context of weather forecasting, predictions of hurricane intensity are strongly dependent on turbulence closure models. 
There are therefore several papers discussing the intricacies of the closure model \citep{foster2009boundary, nolan2009evaluation, smith2010dependence, kepert2012choosing, gopalakrishnan2013study, gopalakrishnan2021improving, zhang2015evaluating, zhang2017impact, chen2021role, chen2022evaluation, chen2022planetary, chen2023performance, romdhani2022characterizing, matak2023role}.
As an initial step toward a more accurate closure, \citet{zhang2011characteristic} and \citet{kepert2012choosing} contend that a dynamical boundary layer model offers a more representative description than a thermodynamical formulation, with the latter continuing to underpin many mesoscale closure models.
Almost all popular dynamical models used in practice are a function of the height of the boundary layer. 
\citet{chen2022planetary} studied the simulation of various features of HBL winds such as the velocity profiles, eddy viscosity and total stress  using four popular mesoscale closure models. It was found that the primary source of variation is due to the differences between the models in the prediction of the HBL height.
In wind engineering too, proposed wind profile models, for example by \citet{vickery2009hurricane}, use HBL height based on a constant eddy viscosity.
It would be desirable to inform these models with hurricane-relevant parameters such as radial distance, and replace the existing reliance on a constant eddy viscosity with more practical parameters such as the friction velocity $u_\star$.

The literature addressing the atmospheric boundary layer (ABL) is grounded in a comparatively more developed framework.
The original proposition by \citet{ekman1905influence} for the constant eddy viscosity scaling has been extended to a commonly accepted turbulent scaling $u_\star/f$ for neutrally stratified flows \citep{blackadar1968asymptotic, tennekes1972first}. 
Since then, there have also been well-accepted modifications to the height scale to include an inversion layer, and for stable and unstable stratification. \citep{zilitinkevich1972determination, pollard1973deepening, kitaigorodskii1988note, kitaigorodskii1988search, joffre2001variability, zilitinkevich2007further}. 
A good review of the historical progress is given in the introduction section of \citet{narasimhan2024analytical}.

This gap in the literature for the HBL can be attributed to the historical inability to perform turbulence-resolving simulations for the HBL. 
The sensitivity analyses in such studies require numerous turbulence-resolving simulations.
However, to simulate the entire hurricane system across its large range of scales, even a single simulation would take several months on a dedicated cluster \citep{bryan2017eddy}.
To circumvent this, \citet{nakanishi2012} proposed a method in which large-eddy simulations (LESs) of the HBL could be conducted in a channel-flow-like setup to resolve a small but representative part of the HBL by including the large-scale rotational effects in the governing equations. 
Modifications to this methodology were proposed by \citet{green2015} and an influential paper by \citet{bryan2017}. 
This has spurred several follow-up papers that target various related applications \citep{worsnop2017spectral,chen2021role, chen2022evaluation, momen2021scrambling,ma2021large,chen2021effect, sabet2022characterizing, richter2025large}.
While this technique is limited to regions outside the eyewall, as it neglects average vertical advection, it now allows for LES of the HBL to be conducted at a reasonable computational cost. 
In particular, one can obtain reliable predictions of the friction velocity and the HBL height, as well as sensitivities to input parameters.

 In this work, we draw inspiration from the literature on neutrally and stably stratified turbulent atmospheric boundary layers, as well as the analytical work on the HBL, to propose a scaling for the HBL height. 
 The scaling is informed and validated using a suite of process-resolving LES of the HBL in the above framework and with available observations.
 
 Section 2 documents the scaling obtained analytically.
 Section 3 compares the proposed formulae with the HBL height obtained directly from the simulations and with wind profiles from field observations of hurricanes. 
 Section 4 establishes relations between standard metrics of the boundary layer height, such as radial and tangential jet heights, and discusses further scaling implications.
 A summary and conclusions are presented in Section 5.

 

\section{Analytical Model for Height Scaling}
\label{sec:analyticalModel}
\subsection{Background}
Most observational profiles of velocity and potential temperature in the HBL exhibit characteristics of stable stratification outside the eyewall and outer rainbands
\citep{zhang2011characteristic,zhang2012observational,ming2015typhoon}.
Stable stratification must hence be accounted for when interpreting observational and mesoscale simulation data.
The primary analytical derivation and validation are performed with simulations and observations possessing stable stratification.
Validation of the formula for neutral stratification is described in the Appendix.

In the presence of stable stratification, it is necessary to account for the competing influence of buoyancy, which acts to inhibit vertical motion and oppose inertial forces. 
Stable stratification is typically quantified by a surface cooling flux $B_s$ \citep{zilitinkevich1972determination}, a constant cooler surface temperature, or a surface cooling rate \citep{narasimhan2024analytical}.
In a hurricane, however, very near the surface, there is unstable stratification due to the presence of warm ocean water below cooler air, and stable stratification above due to latent heat release from cloud formation coupled with a gradual subsidence outside the eyewall. 
This switch in stratification from unstable to stable, therefore, involves complex coupling between surface heat flux, evaporation and condensation, radiation, etc.
Additionally, there is the practical difficulty of obtaining accurate measurements of heat flux near the surface during a hurricane.
Stratification strength is therefore often characterized by the profile of potential temperature alone.
For example, \citet{chen2021framework} nudge to a known potential temperature profile to circumvent modeling the complex physics of radiation and moisture.
We therefore attempt to analytically characterize stratification in a similar manner.

Monin-Obukhov theory (MOST) states that \citep{wyngaard2010turbulence}
\begin{equation}
    \frac{\kappa z}{\theta_\star}\frac{d\theta}{dz} =
    -\frac{\kappa zu_\star}{H}\frac{d\theta}{dz}
    =\phi_h(\zeta) = 1 + 7.8\frac{z-z_0}{L_s}
\end{equation}
where the Obukhov length is
\begin{equation}
    L_s = 
    -\frac{u_\star^3}{B_s}
    = -\frac{u_\star^3\theta_0}{\kappa g H}\ ,
\end{equation}
$B_s$ is the dynamic heat flux and $H$ is the kinematic heat flux.
If we choose a sufficiently large $z$, we can write
\begin{equation}
    -\frac{\kappa zu_\star}{H}\frac{d\theta}{dz}
    =
    -7.8\frac{z\kappa g H}{u_\star^3\theta_0}
    \ .
\end{equation}
Solving for $H$ and hence for $L_s$, we have
\begin{equation}
    L_s = \frac{u_\star}{\kappa  }\sqrt{\frac{7.8\theta_0}{gd\theta/dz}}
    = \frac{u_\star}{N}\frac{\sqrt{7.8}}{\kappa}
    \ ,
\end{equation}
where $N$ is the Brunt-V\"ais\"al\"a frequency.
Thus, the Obukhov length can be expressed roughly as $\sim u_\star/N$, where we assume $N$ is a constant for simplicity. 
\citet{kitaigorodskii1988note} uses a similar characterization and refers to this in his work as ``imposed stable stratification" or ``background stratification".

The time-averaged, linearized HBL equations \citep{kepert2001dynamics, sathia2025analytical} are
\begin{eqnarray}
\label{eq:HBLlinearizedEqn_1}
    \alpha(V_g - v) 
    &=& 
    \frac{d}{dz}(\tau_{xz}) 
    \ ,
    \\
    \label{eq:HBLlinearizedEqn_2}
    \beta u 
    &=& 
    \frac{d}{dz}(\tau_{yz})
    \ ,
\end{eqnarray}
where $u$ and $v$ are the radial and tangential components of the velocity, $\tau_{xz}$ and $\tau_{yz}$ are the components of the stress along these directions, $V_g$ is the gradient wind speed, $f$ is the Coriolis frequency, $n$ is the normalized radial derivative of the gradient wind $-\frac{R}{V_g}\frac{\partial V_g}{\partial r}$ and
\begin{equation}
    \alpha = f + \frac{2V_g}{R} \ , \qquad
    \beta = f + \frac{(1-n)V_g}{R} \ .
\end{equation}
$\alpha$ is twice the absolute angular velocity and $\beta$ is the absolute fluid vorticity \citep{smith2020generalized,sous2013friction}.

A typical approach to determining the boundary-layer height in the ABL relies on a mixing-length argument.
Consider representative length scales
\begin{equation}
    |V_g - v| \sim \tilde{v} 
    \ ,
    \qquad \qquad 
    |u| \sim \tilde{u}
    \ .
\end{equation}
Stresses are modeled using an eddy viscosity $K_m$, which is parameterized as $K_m \sim u_\star l_T$, 
where $u_\star$ is the friction velocity, and $l_T$ is the mixing length, a turbulent length scale representative of the largest eddies within the boundary layer \citep{zilitinkevich2007further}. 
This gives us
\begin{equation}
    \alpha\tilde{v}
    \sim 
    \frac{u_\star l_T}{h^2}\tilde{u}
    \ , 
    \qquad \qquad
    \beta\tilde{u} 
    \sim
    \frac{u_\star l_T}{h^2}\tilde{v}
    \ .
\end{equation}
Typical choices for $l_T$ are $h$ under neutral stratification, and $-u_\star^3/B_s \sim u_\star/N$ under stable stratification \citep{zilitinkevich2007further}.
Substituting and simplifying, we obtain $h\sim u_\star/I$ under neutral stratification and $h\sim u_\star/\sqrt{IN}$ under stable stratification, where $I$ is the inertial frequency \citep{kepert2001dynamics}
\begin{equation}
    I=\sqrt{\alpha \beta} = \sqrt{\left(f+\frac{2V_g}{R}\Big)\Big(f+\frac{(1-n)V_g}{R}\right)}\ .
\end{equation}
Recall that the constant eddy-viscosity scaling for the HBL is $\sqrt{2K/I}$ and that the turbulent ABL scalings are $u_\star/f$ and $u_\star/\sqrt{fN}$.
The formulae obtained above from the mixing-length argument thus seem to be a reasonable, natural extension to the constant eddy viscosity expression.
These, however, are inconsistent with our findings. 
The formulae we propose instead are
\begin{equation}
\label{eq:neutralHBLScaling}
    h = C_R\frac{u_\star}{\beta}
\end{equation}
under neutral stratification, and
\begin{equation}
\label{eq:stratifiedHBLscaling}
    h = C_S\frac{u_\star}{\sqrt{\beta N}}
\end{equation}
under stable stratification.
The following subsection derives the expression under stable stratification. While we do not possess a derivation for the neutral stratification case, the proposed formula is the same as the empirical suggestion by \citet{sous2013friction} for a spin-down flow in a rotating tank. 
It is also provides accurate predictions of the boundary layer height obtained from our LES runs, and is described further in the Appendix.

\subsection{Derivation}
We follow the derivation of \citet{pollard1973deepening} (henceforth P73). 
P73 had derived an expression for the deepening of the upper ocean mixed layer, due to an imposed wind stress, against quiescent, stably stratified ocean water beneath.
We derive a similar expression for the deepening of the stratified HBL due to the gradient wind and compare it to the one in P73. 

Starting from \eqref{eq:HBLlinearizedEqn_1} and~\eqref{eq:HBLlinearizedEqn_2} but retaining the unsteady terms, we have
\begin{eqnarray}
    \frac{du}{dt} + \alpha(V_g - v) &=& \frac{d}{dz}(\tau_{xz})
    \ ,
    \\
    \frac{dv}{dt} + \beta u &=& \frac{d}{dz}(\tau_{yz})
    \ .
\end{eqnarray}
Integrating from $0$ to $h$ to obtain a slab model \citep{kepert2010slab}, the equations become
\begin{eqnarray}
    \frac{d(h\overline{u})}{dt} 
    - u\Big|_h\frac{dh}{dt} 
    + \alpha h(V_g - \overline{v}) 
    &=& 
    \tau_{xz}\Big|_h - \tau_{xz}\Big|_0
    \ ,
    \\
    \frac{d(h\overline{v})}{dt} 
    - v\Big|_h\frac{dh}{dt} 
    + \beta h\overline{u} 
    &=& 
    \tau_{yz}\Big|_h - \tau_{yz}\Big|_0
    \ .
\end{eqnarray}
Here we have used the Leibniz rule
\begin{equation}
    \frac{d}{dt}\int_0^{h(t)}u(t,z)dz = \int_0^{h(t)}\frac{\partial u}{\partial t}dz + u(t,h)\frac{dh}{dt}
    \ .
\end{equation}
At the surface, we use a standard drag parameterization \citep{kepert2010slab}
\begin{eqnarray}
\label{eq:wallStress_x}
    \tau_{xz}\Big|_0 &=& C_D\overline{u}\sqrt{\overline{u}^2 + \overline{v}^2} 
    \ ,
    \\
    \label{eq:wallStress_y}
    \tau_{yz}\Big|_0 &=& C_D\overline{v}\sqrt{\overline{u}^2 + \overline{v}^2} 
    \ .
\end{eqnarray}
We first linearise about $\overline{u} = 0$ and $\overline{v} = V_g$ to obtain
\begin{eqnarray}
\label{eq:wallStress_x_linearized}
    \tau_{xz}\Big|_0 &=& C_DV_g\overline{u} 
    \ ,
    \\
    \label{eq:wallStress_y_linearized}
    \tau_{yz}\Big|_0 &=& C_DV_g^2  + 2C_DV_g(\overline{v} - V_g)
    \ .
\end{eqnarray}
The terms $C_DV_g\overline{u}$ and $2C_DV_g(\overline{v} - V_g)$ correspond to linear Rayleigh damping.
For analytical convenience, we drop the damping terms and model the surface stresses as 
\begin{eqnarray}
\label{eq:wallStressSimple}
    \tau_{xz}\Big|_0 = 0
    \ ,
    \qquad \qquad
    \tau_{yz}\Big|_0 = C_DV_g^2    \ .
\end{eqnarray}
We later integrate numerically, retaining the surface stresses in the form of \eqref{eq:wallStress_x} and~\eqref{eq:wallStress_y} to show that this is a robust approximation for this derivation.

At the top of the HBL, the stress is due to the frictional forces from the momentum exchange at the interface between the HBL and the ambient gradient wind region \citep{haiden2005katabatic}.
We neglect this stress.
\begin{eqnarray}
    \tau_{xz}\Big|_h = 0 \ ,
    \qquad \qquad
    \tau_{yz}\Big|_h = 0
    \ .
\end{eqnarray}
Substituting, we obtain
\begin{eqnarray}
    \label{eq:EkmanIVP_1}
    \frac{d\tilde{u}}{dt} - \alpha\tilde{v} &=&  0
    \ ,
    \\
    \label{eq:EkmanIVP_2}
    \frac{d\tilde{v}}{dt} + \beta\tilde{u}  &=& 
     - C_DV_g^2
     \ .
\end{eqnarray}
where $\tilde{u} = h\overline{u}$ and $\tilde{v} = h(\overline{v} - V_g)$.
Solving with initial conditions $\tilde{u} = 0$, $\tilde{v} = 0$, we get
\begin{eqnarray} 
\label{eq:hutildeSoln}
h\overline{u}(t)
&=& -\sqrt{\frac{\alpha}{\beta}}\frac{C_DV_g^2}{I}
\left(
1 - \cos(It)
\right),
\\
\label{eq:hvtildeSoln}
h(\overline{v} - V_g)(t)
&=& -\frac{C_DV_g^2}{I}
\sin(It)
.
\end{eqnarray}
For the stratified ABL, we have $\alpha = \beta = f$, and these reduce to the same expressions obtained in P73. 

To close the system, we require an additional equation, for which we use the bulk Richardson number
\begin{equation}
\mathrm{Ri}_b = \frac{g}{\theta_0}\frac{d\theta}{dz}\frac{h^2}{(\Delta u)^2}\ .
\end{equation}
Shear is the main driver of turbulent mixing, whereas stable stratification suppresses it through buoyancy. 
When $\mathrm{Ri_b}$ reaches a critical value $\mathrm{Ri_c}$, shear production balances buoyancy destruction. 
Once this balance is attained, turbulent entrainment ceases, and the height of the turbulent boundary layer (HBL) no longer increases.
In the analytical model, following P73, we impose $\mathrm{Ri_b} = \mathrm{Ri_c}$ throughout the growth phase, so that the evolving boundary layer is treated as remaining in this marginally balanced state up to the point where growth terminates.
This closure condition yields
\begin{equation}
\label{eq:hClosure}
    h^2 = \frac{\mathrm{Ri_c}}{N^2}(\overline{u}^2 + (\overline{v} - V_g)^2)\ .
\end{equation}
Squaring and adding \eqref{eq:hutildeSoln} and \eqref{eq:hvtildeSoln}, and substituting in \eqref{eq:hClosure}, we obtain
\begin{equation}
\label{eq:squareAndAdd}
    h^2\overline{u}^2 + h^2(\overline{v} - V_g)^2 =
    \frac{N^2}{\mathrm{Ri_c}}h^4 =  C_D^2V_g^4\left(\frac{1}{\beta^2}(1-\cos(It))^2 + \frac{1}{I^2}\sin(It)^2\right)\ .
\end{equation}

When $It = \pi$, the HBL stops growing and \eqref{eq:squareAndAdd} reaches its maximum value
\begin{equation}
    h = 
    \left(\sqrt{2}\mathrm{Ri_c}^{0.25}
    \right)
    \sqrt{\frac{C_DV_g^2}{\beta N}}
    \ .
\end{equation}
$C_DV_g^2$ can be written equivalently as $u_\star^2$, and so we have
\begin{equation}
\label{eq:stratifiedHBLscaling}
\boxed{
    h = C_S 
    \frac{u_\star}{\sqrt{\beta N}}
    }
\end{equation}
where $C_S$ is a tuning constant. 
We show in Section 3 that, from LES results, $C_S = 1.2$.
Thus, the parameter in the denominator ends up being $\beta$ instead of $I = \sqrt{\alpha\beta}$.

Once again, as a sanity check, we confirm that as $R\to\infty$, we have $\beta\to f$ and \eqref{eq:stratifiedHBLscaling} reduces to $u_\star/\sqrt{fN}$ which is the well accepted expression for the stratified ABL \citep{pollard1973deepening,zilitinkevich2007further}.
\begin{figure} 
\centering \includegraphics[width=0.5\textwidth]{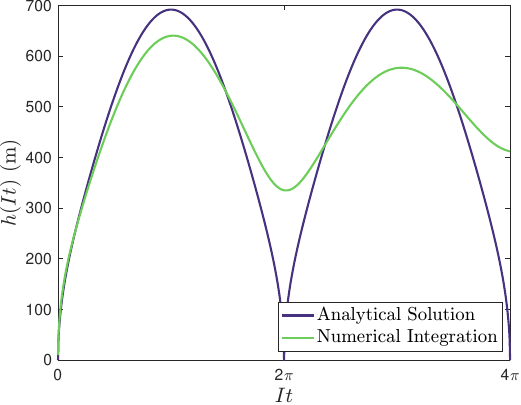}
 \caption{Comparison between analytical solution \eqref{eq:squareAndAdd} and numerical integration of \eqref{eq:numericalInteg1} and~\eqref{eq:numericalInteg2}. Constants used are $f = 10^{-4}$ s$^{-1}$, $N = 1.28\times10^{-2}$ s$^{-1}$, $C_D = 10^{-3}$, $V_g = 45$ ms$^{-1}$, $R = 40$ km, $n = 0.5$, $\tau = 343$ s.}
 \label{fig:numericalIntegration}
\end{figure}
The two main approximations introduced above for analytical tractability are (a) the simplification of the stress terms, going from \eqref{eq:wallStress_x}, \eqref{eq:wallStress_y} to \eqref{eq:wallStressSimple}, and (b) the assumption that the Richardson number remains fixed at its critical value.
To verify that these approximations do not substantially affect the resulting scaling, we numerically integrate the following equations.
\begin{equation}
\label{eq:numericalInteg1}
    \frac{d\overline{u}}{dt} + \alpha (V_g - \overline{v}) = 
    -\frac{\overline{u}}{h}\frac{d\overline{h}}{dt} - \frac{C_D\overline{u}}{h}\sqrt{\overline{u}^2 + \overline{v}^2}
    \ ,
\end{equation}
\begin{equation}
    \label{eq:numericalInteg2}
    \frac{d\overline{v}}{dt} + \beta \overline{u} = 
    -\frac{(\overline{v} - V_g)}{h}\frac{d\overline{h}}{dt} - 
    \frac{C_D\overline{v}}{h}\sqrt{\overline{u}^2 + \overline{v}^2}
    \ ,
\end{equation}
\begin{equation}
    \frac{dh}{dt} = 
    \frac{1}{\tau}
    \left(
    \frac{\mathrm{Ri_c}}{N^2}(\overline{u}^2 + (\overline{v} - V_g)^2) - h
    \right)
    \ .
\end{equation}

Boundary layer growth is modelled using a simple relaxation equation \citep{stull2012introduction}, in which $h \to h_{eq}$, and $h_{eq}$ is defined by \eqref{eq:hClosure}. 
We set the characteristic timescale $\tau = 343$ s, corresponding to $h/u_\star$ diagnosed from the LES for the parameter set used in the numerical integration. 
A representative choice of $\tau$ is sufficient since we seek the peak value of $h$.

We plot two curves in Fig.~\ref{fig:numericalIntegration}. 
The purple curve plots $h$ obtained from~\eqref{eq:squareAndAdd}.
The green curve is obtained from the numerical integration.
The green curve follows, as expected, roughly the same amplitude and frequency as the purple one, albeit with damping. 

We can attempt to combine~\eqref{eq:neutralHBLScaling} and~\eqref{eq:stratifiedHBLscaling} into a single formula.  
Similar to the approach in \citet{zilitinkevich2007further}, we propose
\begin{equation}
    h = \frac{u_\star}{\left(
    \left(\frac{1}{C_R}\beta\right)^4
    +
    \left(\frac{1}{C_S}\sqrt{\beta N}\right)^4
    \right)^{1/4}}\ .
\end{equation}
Validation for this formula against LES results is provided in the Appendix.

\section{Model Validation}
In deriving the scaling in Section~\ref{sec:analyticalModel}, 
we did not specify a precise definition of the boundary layer height. 
We now define the boundary layer height $h$ explicitly as the height at which the turbulent kinematic stress becomes negligible.
Boundary layer height is obtained from the simulations as the height of the first grid point at which the kinematic stress $\sqrt{(-\overline{u'w'} + \tau_{xz}^{SGS})^2 + (-\overline{v'w'} + \tau_{yz}^{SGS})^2}$ achieves a value $< 0.02\ u_\star^2$.

\label{sec:validation}
\subsection{Numerical Simulations}
The governing equations solved in the LES are
\begin{eqnarray}
    \frac{\partial u}{\partial t} + 
(\mathbf{u}\times (\nabla\times\mathbf{u}))_x
&=&
-\frac{\partial \Phi}{\partial x} 
-f(V_g - v)
+   
 \left( 
\frac{\langle u\rangle ^2}{R}
+\frac{\langle v\rangle ^2}{R}
-\frac{V_g^2}{R} \right)
-(\nabla\cdot \tau^{SGS})_x
\ ,
\\
    \frac{\partial v}{\partial t} + 
(\mathbf{u}\times (\nabla\times\mathbf{u}))_y
&=& 
-\frac{\partial \Phi}{\partial y}
-fu
- 
\left( 
\frac{\langle u\rangle \langle v\rangle}{R} 
-
n\langle u \rangle \frac{V_g}{R} 
\right)
-(\nabla\cdot \tau^{SGS})_y
\ ,
\\
\frac{\partial w}{\partial t} + 
(\mathbf{u}\times (\nabla\times\mathbf{u}))_z
&=&
-\frac{\partial \Phi}{\partial z}
+ g\frac{(\theta- \langle \theta \rangle)}{\theta_0}
-(\nabla\cdot \tau^{SGS})_z 
\ ,
\\
\frac{\partial \theta}{\partial t} + 
\mathbf{u}\cdot\nabla\theta
&=&
 \frac{(\theta_r - \langle \theta \rangle)}{\tau_r}
-(\nabla\cdot \pi^{SGS})
\ ,
\end{eqnarray}
where $x$ is the outward radial, $y$ is the tangential and $z$ is the vertical direction.

The primary validation is performed with simulations possessing stable stratification.
For these stratified cases, we use the simulation set \citet{sathia2026database} uploaded to the NHERI DesignSafe Database. 
Details of the numerical simulations are provided in the Appendix.
We use a linearly increasing reference profile $\theta_r = 300 + (5\times 10^ {-3})z$ Kelvin, as described in \citet{chen2021framework}. 
Additional simulations are conducted with neutral and mild stratification. 
These are also described in the SI Appendix.

Note here that the governing equations of the LES include the nonlinear terms, whereas the derivation of the height scales was performed using the linearized form of the equations.
We assume that the nonlinearity contributes only a correction to the solution of the linearized equations and follows the same height scale \citep{kepert2001nonlinear,foster2009boundary,sathia2025analytical}.

\subsection{Observations}
To validate the proposed scaling against observations, one requires not only mean velocity profiles but also the corresponding values of $R$, $n$, and $N$. 
In practice, even obtaining robustly averaged HBL profiles is nontrivial, let alone those conditioned on specific $R$ or $n$.
As a preliminary assessment, we therefore use the composite hurricane profiles reported in \citet{bryan2017} and \citet{chen2021framework}, adopting the values of $R$ and $n$ provided therein.
The vertical shear of the gradient wind, $G_{z}$, is estimated as the slope of a linear fit to $v$ above the wind maximum. 
The gradient wind $G$ is then computed as the mean of $v - G_{z} z$ over the same region. 
These estimates are consistent to leading order with the values reported in the original studies.

The friction velocity is obtained by fitting logarithmic profiles to the near-surface winds (below 50 m), namely $u_{\star,y} \log(z/z_0)/\kappa$ to $v$ and $u_{\star,x} \log(z/z_0)/\kappa$ to $u$, and then forming $u_\star = \sqrt{u_{\star,x}^2 + u_{\star,y}^2}$. 
For all five cases, we find $u_\star \approx u_{\star,y}$.

For the Chen et al. composites, the reported potential temperature profiles correspond to $d\overline{\theta}/dz \approx 5\ \mathrm{K,km^{-1}}$. 
Although $d\overline{\theta}/dz$ is not explicitly available for the Bryan et al. cases, semi-logarithmic plots of $v$ reveal comparable slopes to those in Chen et al., suggesting similar stratification and thus comparable $N$ across all profiles.
We find that a slightly larger $C_S = 1.44$ better fits the data and use this value for the observational analysis.
The parameter values used to estimate $h$ for each profile are summarized in Table \ref{tab:obs}.

\begin{table*}
\caption{List of parameter values and predicted BL height for observations}
\label{tab:obs}
\begin{center}
\begin{tabular}{ccccccccc}
\hline\hline
&$G$ (ms$^{-1}$) & $R$ (km) & $n$ & $f$ ($10^{-4}$ s$^{-1}$) & $G_{z}$ ($10^{-3}$ s$^{-1}$) & $N$ ($10^{-2}$ s$^{-1}$) & $u_\star$ (s$^{-1}$) & $h_{\mathrm{pred}}$ (m)\\
\hline
Chen\_V25 ($\ast$) &
$42$ &
$110$ &
$0.7$ &
$0.5$ &
$0$ &
$1.2$ &
$1.22$ &
$1212$
\\
Chen\_V35 ($\square$) &
$58$ &
$75$ &
$0.7$ &
$0.5$ &
$-4$ &
$1.2$ &
$1.36$ &
$1031$
\\
Chen\_V45 ($\circ$) &
$68$ &
$40$ &
$0.75$ &
$0.5$ &
$-3$ &
$1.2$ &
$1.34$ &
$781$
\\
Bryan\_V40 ($\bullet$) &
$40$ &
$40$ &
$0.8$ &
$0.5$ &
$-2.2$ &
$1.2$ &
$0.93$ &
$749$
\\
Bryan\_V60 ($\triangle$) &
$59$ &
$40$ &
$0.73$ &
$0.5$ &
$-3.3$ &
$1.2$ &
$1.17$ &
$710$
\\
\hline
\end{tabular}
\end{center}
\end{table*}

\begin{figure}[t]
     \centering
\includegraphics[width=0.5\textwidth]{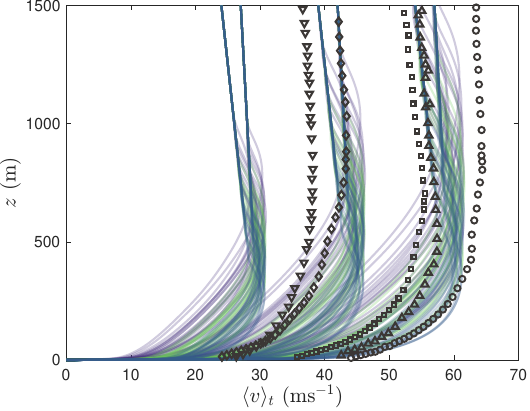}
 \caption{Sensitivity of tangential velocity to input parameters in dimensional form. Purple curves correspond to simulations with $z_0 = 10^{-1}$ m, green to $z_0 = 10^{-2}$ m and blue to $z_0 = 10^{-3}$ m. Symbols correspond to observations (see Table \ref{tab:obs}).}
 \label{fig:stabStratDimSens}
\end{figure}
\subsection{Validation}
\begin{figure}
\centering
 \noindent
\includegraphics[width=0.5\textwidth]{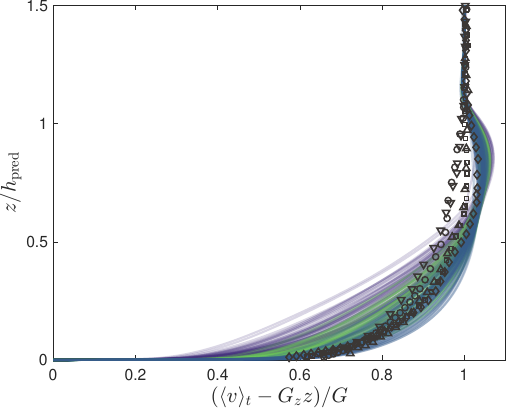}
\caption{Tangential velocity in normalized form.}
 \label{fig:normedCurvesTangential}
\end{figure}

\begin{figure}
\centering \noindent
\includegraphics[width=0.5\textwidth]{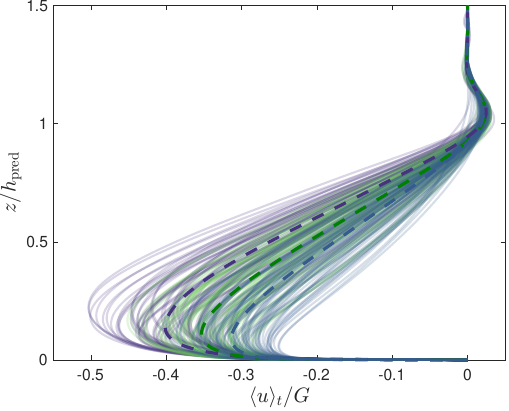}
 \caption{Radial velocity in normalized form. Dashed lines denote the average of the curves of the same color. }
 \label{fig:normedCurvesRadial}
\end{figure}

Figure~\ref{fig:stabStratDimSens} shows the tangential velocity time- and horizontally spatially-averaged (denoted by $\langle\cdot\rangle_t$) from both simulations and observations, demonstrating the dimensional variation to $G$, $R$, $z_0$, $n$ and $f$.
Figure~\ref{fig:normedCurvesTangential} shows the tangential velocity profiles for all 216 LES runs and the observations in normalized form, with the height scaled by $C_Su_\star/\sqrt{\beta N}$.
We see that the LES profiles collapse well under the proposed scaling.
The primary variation near the ground for both the tangential (Fig.~\ref{fig:normedCurvesTangential}) and radial velocities (Fig.~\ref{fig:normedCurvesRadial}) is due to that of $z_0$.
The blue curves are for $z_0 = 10^{-3}$ m, the green curves for $z_0 = 10^{-2}$ m and the purple curves for $z_0 = 10^{-1}$ m.
The remaining variation within a color is due to the variations in $R$ and $G$.
These variations arise from changes in the friction velocity, driven — in order of decreasing influence — by $z_0$, $R$ and $G$.
The parameters $f$, $n$ and $G_{z}$ have a negligible effect on the friction velocity, and therefore collapse almost perfectly into a single profile.


The observations in normalized form also fall within the range of variation of the simulated profiles, and collapse well in the outer region. 
The dimensional V45 profile exhibits a pronounced nose and a non-zero radial velocity $u$ in the outer layer. Both features are characteristic of flow in, or proximate to, the eyewall region—a regime not accounted for in the present derivation. 
Despite this mismatch in underlying assumptions, the profile nevertheless collapses satisfactorily with the others, and the proposed scaling continues to perform well.

\begin{figure}
\centering
\includegraphics[width=0.5\textwidth,clip=true]{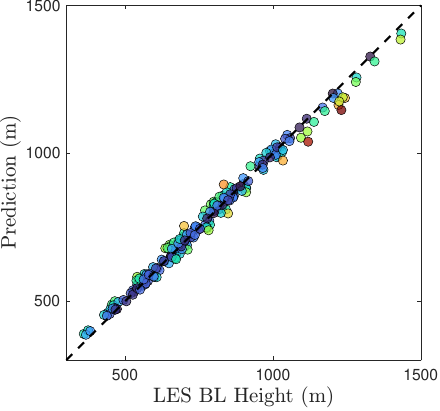}
 \caption{Parity plot comparing LES-derived HBL height against prediction from \eqref{eq:stratifiedHBLscaling}. $R^2 = 0.99$, bias $= 1.04$ m and RMSE $= 23.50$ m. Colors encode error magnitude - blue denotes a small error, and red denotes a large error.}
 \label{fig:parity}
\end{figure}

Figure~\ref{fig:parity} shows a plot of LES-derived HBL height against the prediction from~\eqref{eq:stratifiedHBLscaling}. 
The proposed scaling is highly predictive, with an average relative error of $\sim 2.5\%$ and an average absolute error of $\sim 18$ m.

\section{Discussion}
Having established a robust scaling for the boundary layer depth $h$, defined as the height at which the total stress becomes negligible, we can now express other characteristic heights of the velocity profiles in terms of $h$.

\begin{figure}
\centering
\noindent\includegraphics[width=0.5\textwidth]{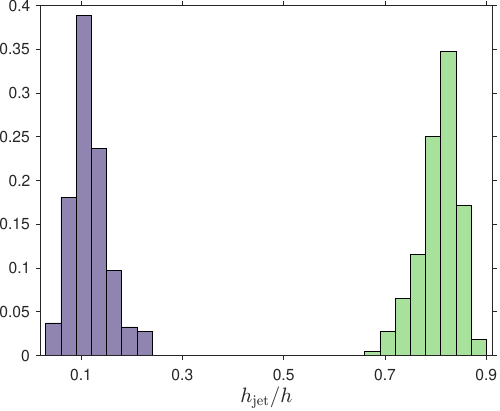}
 \caption{Histograms of radial (purple) and tangential (green) jet peak heights, normalized by the predicted HBL height.}
 \label{fig:histogram}
\end{figure}

The height of peak tangential velocity (from Fig.~\ref{fig:normedCurvesTangential}) and that of the velocity magnitude (not shown) is around $80\%$ of the HBL height, with mild sensitivity in the normalized form to variation in $G$ and $R$.
This finding is consistent with the observational study by \citet{zhang2011characteristic}.
Figure~\ref{fig:normedCurvesRadial} plots the normalized radial velocity variation. 
There is considerable variation near the radial inflow peak, with the peak height varying between $6-20\%$ of the HBL height. 
Figure~\ref{fig:histogram} plots a histogram showing the variation of the height of peak tangential and radial velocities, normalized by the boundary layer height $h_{jet}/h$. 
This ratio, for the tangential velocity peak, lies roughly between 0.65 and 0.85, with a clear peak near 0.8.
The curves collapse better near the HBL height.
The inflow depth, defined as the first grid point at which the velocity profile attains a positive value, varies between $0.86-0.98\ h$.
A similar characterization can also be made for the tangential velocity. 
The depth of the tangential velocity profile, defined as the first grid point above its peak where the profile attains a value less than $G$, varies between $1.08-1.11\ h$.

The proposed scaling also allows us to analyse the overall variation of these quantities to radial distance. 
From \eqref{eq:stratifiedHBLscaling} we have
\begin{equation}
    h \sim \frac{u_\star}{\sqrt{N\left(f + (1-n)\frac{G}{R}\right)}}
    \ .
\end{equation}
Assuming a constant $N$ and $n$ (in particular, which do not vary with $R$) and neglecting $f$, the expression simplifies to
\begin{equation}\label{eq:h_radialScaling}
    h \sim u_\star\sqrt{\frac{R}{G}}
    \ .
\end{equation}

Typically, $u_\star \sim G$. 
$u_\star$ is, as discussed in Section~\ref{sec:validation}, somewhat sensitive to $z_0$ and to a lesser extent to $R$, and is relatively insensitive to all other parameters. 
We assume that $z_0$ stays constant with $R$, and neglect the small dependence of $u_\star$ to $R$. 
Additionally, we substitute the standard assumption $G\sim R^{-n}$ \citep{bryan2017}. 
Substituting these in \eqref{eq:h_radialScaling}, we obtain that the boundary layer height varies with $R$ as 
\begin{equation}
    h \sim R^{\frac{1-n}{2}}
    \ .
\end{equation}
Since the radial and tangential jet heights are simply fractions of the HBL height, this scaling must also hold true for them.
This scaling can also be contrasted against that for neutral stratification where $h \sim u_\star/\beta \sim R$.

We can similarly interpret the variation of eddy viscosity with $R$.
The eddy viscosity typically scales as $K_m\sim u_\star l_T$ where $l_T$ is the mixing length.
Typically, the background stratification (characterized by $u_\star/N$) puts a stronger limit on $l_T$ than the rotation (characterized by $u_\star/\beta$) \citep{zilitinkevich2003effect}.
We therefore have
\begin{equation}
    K_m \sim u_\star^2/N \sim G^2 \sim R^{-2n}\ .
\end{equation}
Thus $K_m$ decreases with radius at a rate that is twice as fast as the gradient wind speed.
This is in contrast to neutral stratification in which $K_m\sim u_\star h \sim G^2/\beta \sim R^{1-n}$ and $K_m$ therefore increases with radius.

Observations of the HBL \citep{stern2009reexamining,zhang2011characteristic, zhang2012observational} typically show that eddy viscosity is large near the eyewall and decreases further away, i.e., decreasing $K$ as $R$ increases. 
This further emphasizes the need to consider a scaling that accounts for stratification when interpreting observational data.

\section{Summary and Conclusions}
In this work, we propose a scaling for the hurricane boundary layer height under neutral and stable stratification. 
These simple scalings draw inspiration from the literature on the analytical treatment of the HBL and on stratified atmospheric boundary layers.
We find that the HBL scales as $h = C_Ru_\star/\beta$ under neutral stratification, and $C_Su_\star/\sqrt{\beta N}$ under stable stratification.

We conducted a comprehensive set of turbulence-resolving HBL simulations spanning physically realistic ranges of input parameters to validate the proposed scaling. 
The LES results suggest that $C_R = 0.58$ and $C_S = 1.2$.
With these values, the formulae predict the LES-derived HBL heights to within $\sim18$ m on average.

The proposed formulae accurately collapse the LES-derived profiles for the nonlinear equations and available observations.
This also enables the characterization of other height scales, such as the height of peak wind speed and the depth of the inflow layer.
We find that the inflow jet peak occurs at $~6-20\%$ of $h$ and the tangential jet peak occurs at $~65-85\%$ of $h$. One can further obtain scaling relationships for the HBL height and the eddy viscosity in terms of input parameters such as $R$.

The proposed scalings provide a quantitative basis for interpreting observational measurements of hurricane boundary layers, where the depth is often difficult to define consistently due to limited vertical resolution and differing diagnostic criteria.
By relating boundary layer height directly to measurable quantities such as surface stress, gradient wind speed, and stratification, the scalings enable a consistent normalization of observed velocity profiles across storms and radial locations, and offer a more practical method for inferring boundary layer depth.



The results also have direct implications for HBL parameterizations in mesoscale and operational models. 
In such models, boundary layer depth and mixing lengths are typically empirically tuned; the proposed formulae provide physically motivated constraints that parameterizations must satisfy. 
The scalings provide a basis for prescribing mean wind profiles, eddy viscosity distributions, vertical momentum transport, and stratification effects.

Finally, these findings are significant for wind engineering and coastal resilience applications, where accurate specification of mean wind profiles and turbulence characteristics is essential. 
The predicted boundary layer height and associated velocity structure are used to define inflow conditions in turbulent inflow generators for studying wind-induced structural response, including tall buildings and wind turbines. 
The simple yet powerful scalings will help decrease uncertainty in simulated wind structure and intensity in reduced-order and surrogate models used in wind energy assessment and coastal hazard analysis.

More broadly, these results unify the roles of rotation and stratification in hurricane boundary layer structure, and connect theory, simulations, and observations across a wide range of conditions.

\clearpage
\acknowledgments

This research was funded by the National Institute of Standards and Technology (\url{ror.org/05xpvk416}) under Grant No. 70NANB22H057. This work also used the Anvil supercomputer at Purdue University through allocation ATM180022 from the Advanced Cyberinfrastructure Coordination Ecosystem: Services \& Support (ACCESS) program, National Science Foundation which supports grants \#2138259, \#2138286, \#2138307, \#2137603, and \#2138296.

%
%
\datastatement

The database of stably stratified simulations is uploaded to the NSF NHERI DesignSafe repository. Profiles of velocity and second-order moments are present therein.
Scripts and other datasets generated as part of this study are available from the corresponding author upon reasonable request.

%






%



\appendix[A]
\appendixtitle{Governing Equations and Numerical Setup}
The equations solved in the large-eddy simulations (LES) are
\begin{eqnarray}
    \frac{\partial u}{\partial t} + 
v\left(\frac{\partial u}{\partial y} -
\frac{\partial v}{\partial x} \right) +
w\left(\frac{\partial u}{\partial z} -
\frac{\partial w}{\partial x} \right)
&=&
-\frac{\partial \Phi}{\partial x} 
-f(V_g - v)+   
{\color{blue} \left( 
\frac{\langle u\rangle ^2}{R}
+\frac{\langle v\rangle ^2}{R}
-\frac{V_g^2}{R} \right)}
-(\nabla\cdot \tau^{SGS})_x
\ , \ \ \ \ \ \ \ \
\\
    \frac{\partial v}{\partial t} + 
u\left(\frac{\partial v}{\partial x} -
\frac{\partial u}{\partial y} \right) +
w\left(\frac{\partial v}{\partial z} -
\frac{\partial w}{\partial y} \right)
&=& 
-\frac{\partial \Phi}{\partial y}
-fu- 
{\color{blue} \left( 
\frac{\langle u\rangle \langle v\rangle}{R} 
-
n\langle u \rangle \frac{V_g}{R} 
\right)}
-(\nabla\cdot \tau^{SGS})_y
\ ,
\\
\frac{\partial w}{\partial t} + 
u\left(\frac{\partial w}{\partial x} -
\frac{\partial u}{\partial z} \right) +
v\left(\frac{\partial w}{\partial y} -
\frac{\partial v}{\partial z} \right)
&=&
-\frac{\partial \Phi}{\partial z}
+ g\frac{(\theta- \langle \theta \rangle)}{\theta_0}
-(\nabla\cdot \tau^{SGS})_z 
\ ,
\\
\frac{\partial \theta}{\partial t} + 
u\frac{\partial \theta}{\partial x} +
v\frac{\partial \theta}{\partial y} +
w\frac{\partial \theta}{\partial z}
&=&
 \frac{(\theta_r - \langle \theta \rangle)}{\tau_r}
-(\nabla\cdot \pi^{SGS})
\ ,
\end{eqnarray}

where $t$ is time, $x$ is the cross-streamwise or radially outward direction, $y$ is the streamwise or tangential direction, and $z$ is the vertical direction. 
$u$ is the filtered radial velocity component, $v$ is the filtered tangential velocity component, and $w$ is the filtered vertical velocity component. 
We use the Boussinesq approximation to account for buoyancy. $\theta$ is the potential temperature, $\theta_0$ is a constant reference temperature, which we set to be $300$ K and $g = 9.81$ ms$^{-2}$. 
$\tau^{SGS}$ is the subgrid-scale (SGS) tensor for velocity, and $\pi^{SGS}$ the SGS vector for potential temperature. 
$\Phi$ is a modified filtered pressure field, namely $\Phi = \frac{p}{\rho} + \frac{1}{3}\tau_{ii}^{SGS} + \frac{1}{2}u_iu_i $ where $\rho$ is a reference constant density and $p$ is the pressure deviation from the mean pressure field imposed from the hydrostatic balance and geostrophic forcing.  $f$ is the Coriolis frequency.
$\langle\cdot\rangle$ denotes a horizontal spatial average. 
The terms in black represent the standard equations of the atmospheric boundary layer, while the terms in blue represent the corrections introduced to account for the large-scale structure of the hurricane. 

The equations are solved with the convective terms in their rotational form \citep{orszag1975numerical} to ensure the conservation of energy in the inviscid limit. 
We use the same governing equations as in \citet{momen2021scrambling}, but with horizontally-averaged terms so as to prevent the growth and transport of spurious oscillations \citep{nakanishi2012,bryan2017}. 
As in these previous works, we assume that the domain is situated sufficiently far from the hurricane eyewall so that average vertical convection is negligible, and that the horizontal domain extent is small compared to $R$.

Following \citet{chen2021framework}, we add a nudging term for potential temperature. 
The role of the potential temperature is simply to modulate the velocity to its correct value; we do not seek to solve for the potential temperature directly, as this would involve the explicit resolution of complex physics, such as radiation.
The reference potential temperature $\theta_r$ we specify as a linear function $300 + 0.005z$ \citep{zhang2011characteristic,chen2021framework}.
$\tau_r$ is a nudging time scale, which we choose to be 1 minute, following \citet{bryan2017}. 
We also ran additional simulations increasing and decreasing $\tau_r$ by a factor of 5, and did not see any change.
We do not explicitly solve for moisture. 
Its effects could be incorporated by interpreting $\theta$ as the virtual potential temperature rather than just the potential temperature. 

The lateral boundary conditions are periodic for all variables. 
A free-lid boundary condition $\frac{\partial u}{\partial z} = \frac{\partial v}{\partial z} = w = \frac{\partial \theta}{\partial z} = 0$ is applied at the top boundary. 
At the bottom boundary, the vertical velocity is set to zero $w = 0$, and $\theta = 302$ Kelvin is set following \citet{chen2021framework}.
Boundary conditions for the horizontal components and the evaluation of tangential SGS stresses are based on an algebraic wall-layer model based on the equilibrium logarithmic law assumption \citep{Chester2007}. 
It is generally well-accepted that the log-law exists for hurricanes near the surface and sufficiently away from the eye.
The roughness length $z_0$ is varied in the simulations to model ocean and land.

\subsection{High-fidelity Numerical Database}
\label{sec:high-fidelity-dataset} 
The gradient wind $V_g = G + G_{z}z$ drives the flow, analogous to the geostrophic wind in the Ekman layer. 
The domain is centered at distance $R$ from the hurricane centre. 
$n$ is the magnitude of the non-dimensional radial derivative of gradient wind $-\frac{dV_g}{dr}\frac{R}{V_g}$, treated as a constant. 
Together with $f$ and $z_0$, these comprises 6 input parameters. 

Based on preliminary sensitivity analyses, we found sensitivity to $V_g$, $R$ and $z_0$ greater and hence assigned them three values each. 
$n$, $G_{z}$ and $f$ are assigned two values each.
Based on the physical range of these parameters and their sensitivity, we use the values listed in Table \ref{tab:Database}.
These comprise 216 simulations. 
The simulation data and averages of the first- and second-order moments have been uploaded to the NSF DesignSafe Data Depot \citep{sathia2026database}.

\begin{table}
    \centering
    \caption{List of parameters for database with stratification}
    \begin{tabular}{cc}
         \textbf{Parameter}& \textbf{Chosen Values}\\
         \hline \\
         $V_g$ (ms$^{-1}$)& 30, 45, 60\\
         $n$& 0.25, 0.5\\
         $G_{z}$ (s$^{-1}$) & -0.02,-0.04\\
         $f$ (s$^{-1}$)& $5\times 10^{-5}$, $1\times 10^{-4}$\\
         $R$ (km)& 40, 80, 120\\
         $z_0$ (m)& 1.00e-03, 1.00e-02, 1.00e-01\\
    \end{tabular}
    \label{tab:Database}
\end{table}

\subsection{Numerical Setup}
\label{sec:numerical_setup}
Simulations are performed in a regular rectangular domain of size $(2\pi\times2\pi\times 2.5)\times 1000$ km using a grid of $256\times256\times512$ collocation nodes in the radial, tangential and vertical directions respectively.
The grid size is consistent with the recommendation in \citet{chen2021framework}.

The LES algorithm was initially developed by Albertson and Parlange \citep{albertson1999natural, albertson1999surface}. 
Equations are solved in strong form on a regular domain, and a pseudo-spectral collocation approach \citep{orszag1969numerical, orszag1970transform} based on truncated Fourier expansions is used in the $x,\ y$ coordinate directions. 
A second-order accurate centered finite differences scheme is adopted in the vertical direction using a staggered grid; the variables $u,\ v,\ \Phi$ are stored at $(j+1/2)\Delta z$, with $j=1,nz$.
Nonlinear terms are fully dealiased via the $3/2$ rule, to avoid piling up energy in the high wavenumber range \citep{kravchenko1997effect, canuto2006spectral}.
SGS stresses in the bulk of the flow are parameterized using the scale-dependent Lagrangian dynamic Smagorinsky model \citep{bou2005scale}. 

Over the past two decades, this solver has been used to develop a series of algebraic SGS closure models for the bulk of turbulent flows \citep{meneveau1996lagrangian, porte2000scale, porte2004scale, bou2005scale, lu2010modulated}, wall-layer models \citep{hultmark2013new}, and immersed-boundary methods to accurately represent solid-gas interfaces \citep{tseng2006modeling, chester2007modeling, fang2011towards, li2016impact}. 
It has also been extensively validated against field and laboratory measurements and used to gain insight into a range of applications involving different flow phenomena, spanning from atmospheric boundary layer flow over flat surfaces to flow over urban areas and forests \citep{tseng2006modeling, bou2005scale, bou2009effects, fang2011towards, shah2014very, pan2014strong, fang2015large, anderson2015numerical, giometto2016spatial, pan2016estimating, giometto2017direct, giometto2017effects}. 
In the context of the hurricane boundary layer (HBL), the solver has been used by \citet{momen2021scrambling} and \citet{sabet2022characterizing} where it has also been validated against observational data.

Simulations are initiated using the uniform fields $V_g$ and $\theta_r$ plus random noise. 
Flow is assumed to be in a fully rough aerodynamic regime, and viscous stresses can be safely neglected.
The turbulent Prandtl number is set to $Pr = 0.7$ (although this has a negligible effect since we are using potential temperature nudging).
A sponge layer with a Rayleigh damping coefficient of 0.01 s$^{-1}$ is used above 2000 m for velocity and temperature.

Time integration is performed via a fully explicit second-order accurate Adams-Bashforth scheme, and a fractional step method is adopted to compute the pressure field \citep{chorin1968numerical, kim1985application}. 
A time-step based on $CFL = 0.075$ is used.
Simulations are run for $1.5T_I$ where $T_I = 2\pi/I$ is the inertial period and $I$ is the inertial frequency \citep{kepert2001dynamics}. 
1000 instantaneous velocity and potential temperature snapshots are collected (skipping the first $0.5T_I$ to achieve statistical stationarity). 

\appendix[B]
\appendixtitle{Validation of proposed formula for neutral and mildly stable stratification}

In addition to the stably stratified simulations \citet{sathia2026database}, we run simulations with neutral and mildly stable stratification. 
These are conducted with the aim of validating the proposed scaling for neutral stratification and to propose a formula that connects the two scalings for mildly stable stratification.

\subsection*{Simulations}

We run cases with small $n = 0,\ 0.25$ for which the total stress profile the Nieuwstadt profile $(1-z/h)^{1.5}$.
For mild and neutral stratification, the stresses follow this profile closely up to $z/h \approx 0.8$, above which they approach zero very slowly.
These simulations, therefore, require a much taller domain, and so the remaining parameters were chosen such that the predicted HBL height is relatively small.
Additionally, this means that the criterion to obtain HBL height used for the stratified simulations (i.e. $2\%$ of $u_\star^2$) does not work well. 
Instead we use the approach described in \citet{narasimhan2024analytical}, where we assume the Nieuwstadt profile holds for the stress, and fit a straight line to $(1-\tau_{\mathrm{total}}/u_\star^2)^{1/1.5}$. 
The HBL height is then obtained as the inverse of the slope.
We verified that for the stratified cases with small $n = 0.25$, this criterion works equally well.

We ran eight simulations with neutral stratification and eight simulations varying the stratification strength $N$. 
The values of $N$ are chosen such that they are in geometric progression from $10^{-4}$ to $10^{-2}\ \mathrm{s}^{-1}$.
A summary listing the various simulations conducted is given in Table~\ref{tab:t1}.
All the numerical details are the same as for simulations with stable stratification \cite{sathia2026database}.

\begin{table}
\caption{List of neutrally and mildly stably stratified LES runs}
\label{tab:t1}
\begin{center}
\begin{tabular}{ccccccccc}
\hline\hline
&$G$ (ms$^{-1}$) & $R$ (km) & $z_0$ (m) & $n$ & $f$ (s$^{-1}$) & $G_{z}$ ($10^{-3}$ s$^{-1}$) & $N$ (s$^{-1}$) & Total Runs\\
\hline
Set A &
$45$, $60$ &
$30$, $35$ &
$0.001$ &
$0$, $0.25$ &
$10^{-4}$ &
$-2$ &
- &
$8$
\\
Set B &
$45$ &
$35$ &
$0.001$ &
$0$ &
$10^{-4}$ &
$-2$ &
\begin{tabular}{@{}c@{}}
$10\hat{\phantom{x}}\{ -4.00,   -3.71,   -3.43, -3.14, $
\\ 
$ -2.86,   -2.57,   -2.29,   -2.00 \}$
\end{tabular} &
$8$ 
\\
\hline
\end{tabular}
\end{center}
\end{table}

\begin{figure}
 \centering
\includegraphics[scale=1]{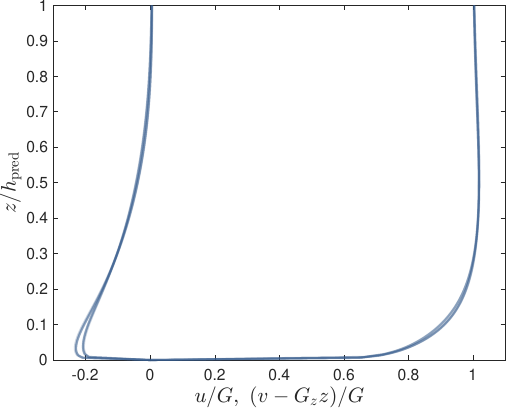}
\caption{Tangential and radial velocity in normalized form for neutrally stratified simulations (Set A of Table \ref{tab:t1})}
\label{fig:app_neutNorm.eps}
\end{figure}

\begin{figure}
\centering
\includegraphics[scale=0.5]{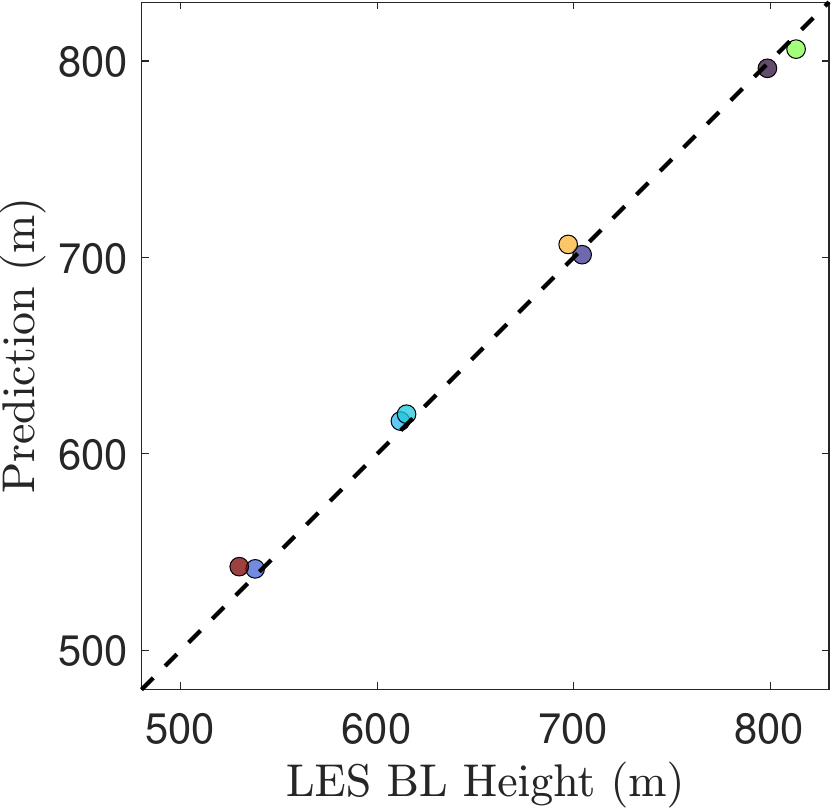}
 \caption{Parity plot comparing LES-derived HBL height against prediction $C_Ru_\star/\beta$. $R^2 = 0.99$, bias $= 2.93$ m and RMSE $= 6.92$ m.}
 \label{fig:app_parity}
\end{figure}

\begin{figure}
\centering
\includegraphics[scale=1]{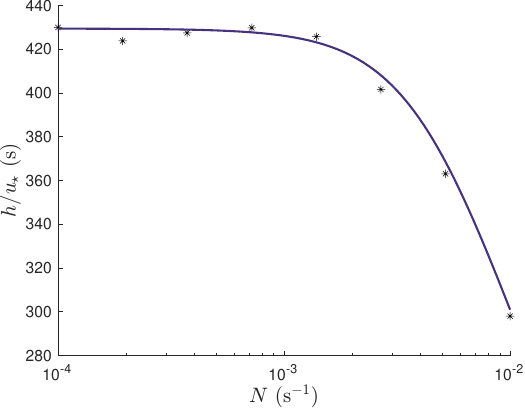}
 \caption{LES-derived (symbols) and predicted (line) HBL heights for mildly stably stratified simulations.}
 \label{fig:app_mildStrat}
 \end{figure}

\subsection*{Validation}
\citet{sous2013friction} had performed experiments of spin-up and spin-down in a rotating tank and suggested a revised scaling of the boundary layer height. 
They recommended $u_\star/\beta$ as a simple empirical modification to account for the fluid rotation, since the fluid feels the absolute vorticity $\beta$ instead of the wall vorticity $f$. 
Figure~\ref{fig:app_neutNorm.eps} shows the profiles for the 8 curves of Set A in Table \ref{tab:t1} in normalized form with the height scaled by the prediction $h_{\mathrm{pred}} = C_Ru_\star/\beta$, with $C_R = 0.58$.
Figure~\ref{fig:app_parity} shows a parity plot of the simulations with neutral stratification against the predictions.
We see from these figures that the predicted height scale works well, with an average error of $6\%$.

To fit a curve that connects the neutrally and stably stratified scalings, we use the approach described in \citet{zilitinkevich2007further}.

\begin{equation}
    \label{eq:mildStrat}
    \left(\frac{1}{h}\right)^p
    =
    \left(\frac{\beta}{C_Ru_\star}\right)^p 
    + 
    \left(\frac{\sqrt{\beta N}}{C_Su_\star}\right)^p
\end{equation}
Figure \ref{fig:app_mildStrat} plots with symbols the boundary layer heights for various values of $N$ obtained from the simulation Set B in Table \ref{tab:t1}.
The purple line is the prediction from \eqref{eq:mildStrat}, using $p = 4$.
While \citet{zilitinkevich2007further} suggested $p = 2$ for the ABL, the figure indicates that a larger value, $p = 4$ provides a better fit. This is possibly because $f$ is much smaller than typical $N$ whereas $\beta$ is more comparable to $N$.

\bibliographystyle{bst}
\bibliography{references}

\end{document}